\documentclass{article}
\usepackage{spconf,amsmath,graphicx}
\usepackage[utf8]{inputenc} 
\usepackage[T1]{fontenc}    
\usepackage{amssymb}
\usepackage{array}
\usepackage{multirow}
\usepackage{array}

\newcolumntype{U}[1]{>{\centering\arraybackslash}p{#1}}

\title{Bidirectional Quaternion Long-Short Term Memory \\ Recurrent Neural Networks for Speech Recognition}
\name{ Titouan Parcollet$^{1,3}$, Mohamed Morchid$^1$, Georges Linarès$^1$, and Renato De Mori$^{1,2}$ }

\address{$^1$Université d'Avignon, LIA, France\\
  $^2$McGill University, Montréal, Canada\\
  $^3$Orkis, Aix en provence, France\\
 }

\begin{document}
%
\maketitle
\begin{abstract}

Recurrent neural networks (RNN) are at the core of modern automatic speech recognition (ASR) systems. In particular, long-short term memory (LSTM) recurrent neural networks have achieved state-of-the-art results in many speech recognition tasks, due to their efficient representation of long and short term dependencies in sequences of inter-dependent features. Nonetheless, internal dependencies within the element composing multidimensional features are weakly considered by traditional real-valued representations. We propose a novel quaternion long-short term memory (QLSTM) recurrent neural network that takes into account both the external relations between the features composing a sequence, and these internal latent structural dependencies with the quaternion algebra. QLSTMs are compared to LSTMs during a memory copy-task and a realistic application of speech recognition on the Wall Street Journal (WSJ) dataset. QLSTM reaches better performances during the two experiments with up to $2.8$ times less learning parameters, leading to a more expressive representation of the information.

\end{abstract}
%

%
%
\section{Introduction}
\label{sec:intro}

During the last decade, deep neural networks (DNN) have encountered a wide success in numerous domain applications. In particular, automatic speech recognition systems (ASR) performances have been remarkably improved with the emergence of DNNs. Among them, recurrent neural networks \cite{medsker2001recurrent} (RNN) have been shown to effectively encode input sequences, increasing the accuracy of neural network based ASR systems \cite{Povey_ASRU2011}. Nonetheless, vanilla RNNs suffer from vanishing/exploding issues~\cite{pascanu2013difficulty}, or the lack of a memory mechanism to remember patterns in very-long or short sequences. These problems have been alleviated by the introduction of  long-short term memory (LSTM) RNN \cite{greff2017lstm} with gates mechanism that allows the model to update or forget information in memory cells, and to select the content cell state to expose in a network hidden state. LSTMs have reached state-of-the art performances in many benchmarks \cite{greff2017lstm,graves2013hybrid}, and are widely employed in recent ASR models, with the almost unchanged acoustic input features used in previous systems.


Traditional ASR systems rely on multidimensional acoustic features such as the Mel filter bank energies alongside with the first, and second order time derivatives to characterize time-frames that compose the signal sequence. Considering that these components describe three different views of the same element, neural networks have to learn both the internal relations that exist within these views, and external or global dependencies that exist between the time-frames. Such concerns are partially addressed by increasing the learning capacity of neural network architectures. Nonetheless, even with a huge set of free parameters, it is not certain that both local and global dependencies are properly represented. To address this problem, new quaternion-valued neural networks, based on a high-dimensional algebra, are proposed in this paper.

Quaternions are hyper-complex numbers that contain a real and three separate imaginary components, fitting perfectly to three and four dimensional feature vectors, such as for image processing and robot kinematics \cite{sangwine1996fourier,aspragathos1998comparative}. The idea of bundling groups of numbers into separate entities is also exploited by the recent capsule network \cite{hinton2017capsule}. With quaternion numbers, LSTMs are conceived to encode latent
inter-dependencies between groups of input features during
the learning process with less parameters than real-valued LSTMs, by taking advantage of the use of the quaternion \textit{Hamilton product} as the counterpart of the dot product. Early applications of quaternion-valued backpropagation algorithms \cite{arena1994neural,arena1997multilayer} have efficiently shown that quaternion neural networks can approximate quaternion-valued functions. More recently, neural networks of hyper-complex numbers have received an increasing attention, and some efforts have shown promising results in different applications. In particular, a deep quaternion network \cite{parcollet2016quaternion, parcollet2017quaternion}, a deep quaternion convolutional network \cite{chase2017quat,parcollet2018qcnn}, or a quaternion recurrent neural network \cite{parcollet2018qrnn} have been successfully employed for challenging tasks such as images, speech and language processing. For speech recognition, in \cite{parcollet2018qcnn}, quaternions with only three internal features have been used to encode input speech. An additional internal feature is proposed in this paper to obtain a richer representation with the same number of model parameters. 
 
Based on all the above considerations, the contributions of this paper can be summarized as follows: 1) The introduction of a novel model, called bidirectional quaternion long-short term memory neural network (QLSTM)\footnote{Code is available at https://github.com/Orkis-Research/Pytorch-Quaternion-Neural-Networks}, that avoids known RNN problems also present in quaternion RNNs, and shows that QLSTMs achieve top of the line results on speech recognition; 2) The introduction of a novel input quaternion that integrates four views of speech time frames. The model is first evaluated on a synthetic memory copy-task to ensure that the introduction of quaternion into the LSTM model does not alter the basic properties of RNNs. Then, QLSTMs are compared to real-valued LSTMs on a realistic speech recognition task with the Wall Street Journal (WSJ) dataset. The reported results show that the QLSTM outperforms the LSTM in both tasks with a higher long-memory capability on the memory task, a better generalization performance with better word error rates (WER), and a maximum reduction of the number of neural paramaters of $2.8$ times compared to real-valued LSTM.

%
%
\section{Quaternion Algebra}
The quaternion algebra $\mathbb{H}$ defines operations between quaternion numbers. A quaternion Q is an extension of a complex number defined in a four dimensional space as:
\begin{align}
Q = r1 + x\textbf{i} + y\textbf{j} + z\textbf{k},
\end{align}
where $r$, $x$, $y$, and $z$ are real numbers, and $1$, \textbf{i}, \textbf{j}, and \textbf{k} are the quaternion unit basis. In a quaternion, $r$ is the real part, while $x\textbf{i}+y\textbf{j}+z\textbf{k}$ with $\textbf{i}^2=\textbf{j}^2=\textbf{k}^2=\textbf{i}\textbf{j}\textbf{k}=-1$ is the imaginary part, or the vector part. 
Such a definition can be used to describe spatial rotations. The {\em Hamilton product} $\otimes$ between two quaternions $Q_1$ and $Q_2$ is computed as follows: 
\begin{align}
Q_1 \otimes Q_2=&(r_1r_2-x_1x_2-y_1y_2-z_1z_2)+\nonumber \\
			&(r_1x_2+x_1r_2+y_1z_2-z_1y_2) \boldsymbol i+\nonumber \\
            &(r_1y_2-x_1z_2+y_1r_2+z_1x_2) \boldsymbol j+\nonumber \\
            &(r_1z_2+x_1y_2-y_1x_2+z_1r_2) \boldsymbol k.
\label{eq:hamilton}
\end{align}
The {\em Hamilton product} is used in QLSTMs to perform transformations of vectors representing quaternions, as well as scaling and interpolation between two rotations following a geodesic over a sphere in the $\mathbb{R}^3$ space as shown in~\cite{minemoto2017feed}.

%
%
\section{Quaternion long-short term memory neural networks}
\label{sec:format}

Based on the quaternion algebra and with the previously described motivations, we introduce the quaternion long-short term memory (QLSTM) recurrent neural network. In a quaternion dense layer, all parameters are quaternions, including inputs, outputs, weights and biases. The quaternion algebra is ensured by manipulating matrices of real numbers \cite{parcollet2018qcnn} to reconstruct the \textit{Hamilton product} from quaternion algebra. Consequently, for each input vector of size $N$, output vector of size $M$, dimensions are split in four parts: the first one equals to $r$, the second to $x$\textbf{i}, the third one is $y$\textbf{j}, and the last one equals to $z$\textbf{k}. The inference process of a fully-connected layer is defined in the real-valued space by the dot product between an input vector and a real-valued $M \times N$ weight matrix. In a QLSTM, this operation is replaced with the \textit{Hamilton product} '$\otimes$' (Eq. \ref{eq:hamilton}) with quaternion-valued matrices (\textit{i.e.} each entry in the weight matrix is a quaternion). 

Gates are core components of the memory of LSTMs. Based on \cite{danihelka2016associative}, we propose to extend this mechanism to quaternion numbers. Therefore, the gate action is characterized by an independent modification of each component of the quaternion-valued signal following a component-wise product (\textit{i.e}. in a \textit{split} fashion \cite{xu2017learning}) with the quaternion-valued gate potential. Let $f_t$,$i_t$, $o_t$, $c_t$, and $h_t$ be the forget, input, output gates, cell states and the hidden state of a LSTM cell at time-step $t$. QLSTM equations can be derived as:

\begin{align}
    f_t =& \sigma(W_{f} \otimes x_t + R_{f} \otimes h_{t-1} + b_f),\\
    i_t =& \sigma(W_{i} \otimes x_t + R_{i} \otimes h_{t-1} + b_i),\\
    c_t =& f_t\times c_{t-1} + i_t\times \alpha(W_{c}x_t + R_{c}h_{t-1} + b_c),\\
    o_t =& \sigma(W_{o} \otimes x_t + R_{o} \otimes h_{t-1} + b_o),\\
    h_t =& o_t \times \alpha(c_t),
\end{align}
with $\sigma$ and $\alpha$ the sigmoid and tanh \textit{quaternion split activations} \cite{xu2017learning,parcollet2016quaternion,parcollet2017deep,arena1997multilayer}. The quaternion weight and bias matrices are initialized following the proposal of \cite{parcollet2018qrnn}. Quaternion bidirectional connections are equivalent to real-valued ones \cite{graves2014towards}. Consequently, past and future contexts are added together component-wise at each time-step. The full backpropagtion of quaternion-valued recurrent neural network can be found in \cite{parcollet2018qrnn}.

%
%
\section{Experiments}
\label{sec:pagestyle}

This section provides the results for QLSTM and LSTM on the synthetic memory copy-task (Section \ref{subsec:copytask}), and a description of the quaternion acoustic features (Section \ref{subsec:acc}) that are used as inputs during the realistic speech recognition experiment with the Wall Street Journal (WSJ) corpus (Section \ref{subsec:wsj}).

%
%
\subsection{Synthetic memory copy-task as a sanity check}
\label{subsec:copytask}

The copy task originally introduced by \cite{hochreiter1997long} is a synthetic test that highlights how RNN based models manage the long-term memory. This characteristic makes the copy task a powerful benchmark to demonstrate that a recurrent model can learn long-term dependencies. It consists of an input sequence of a length $L$, composed of $S$ different symbols followed by a sequence of time-lags or \textit{blanks} of size $T$, and ended by a delimiter that announces the beginning of the copy operation (after which the initial input sequence should be progressively reconstructed at the output). In this paper, the copy-task is used as a sanity check to ensure that the introduction of quaternions on LSTM models does not harm the basic memorization abilities of the LSTM. The QLSTM is composed of $8$K parameters with one hidden layer of size $20$, while the LSTM is made of $8.2$K parameters with an hidden dimension of $40$ neurons. It is worth underlying that due to the nature of the task, the output layer of the QLSTM is real-valued. Indeed, $9$ symbols are one-hot encoded ($S=0,...,7$ for the sequence and $8$ for the \textit{blank}) and can not be split in four components. Different values of $T=10,50,100$ are investigated alongside with a fixed sequence size of $L=10$. Models are trained with the Adam optimizer, with an initial learning rate $\lambda=5\cdot10^{-3}$, and without employing any regularization methods. The training is performed on $2,000$ epochs with the cross-entropy used as the loss function. At each epoch, models are fed with a batch of $10$ randomly generated sequences. 

\begin{figure}[h!]
 \begin{center}
 \scalebox{0.48}{
\includegraphics[width=\textwidth]{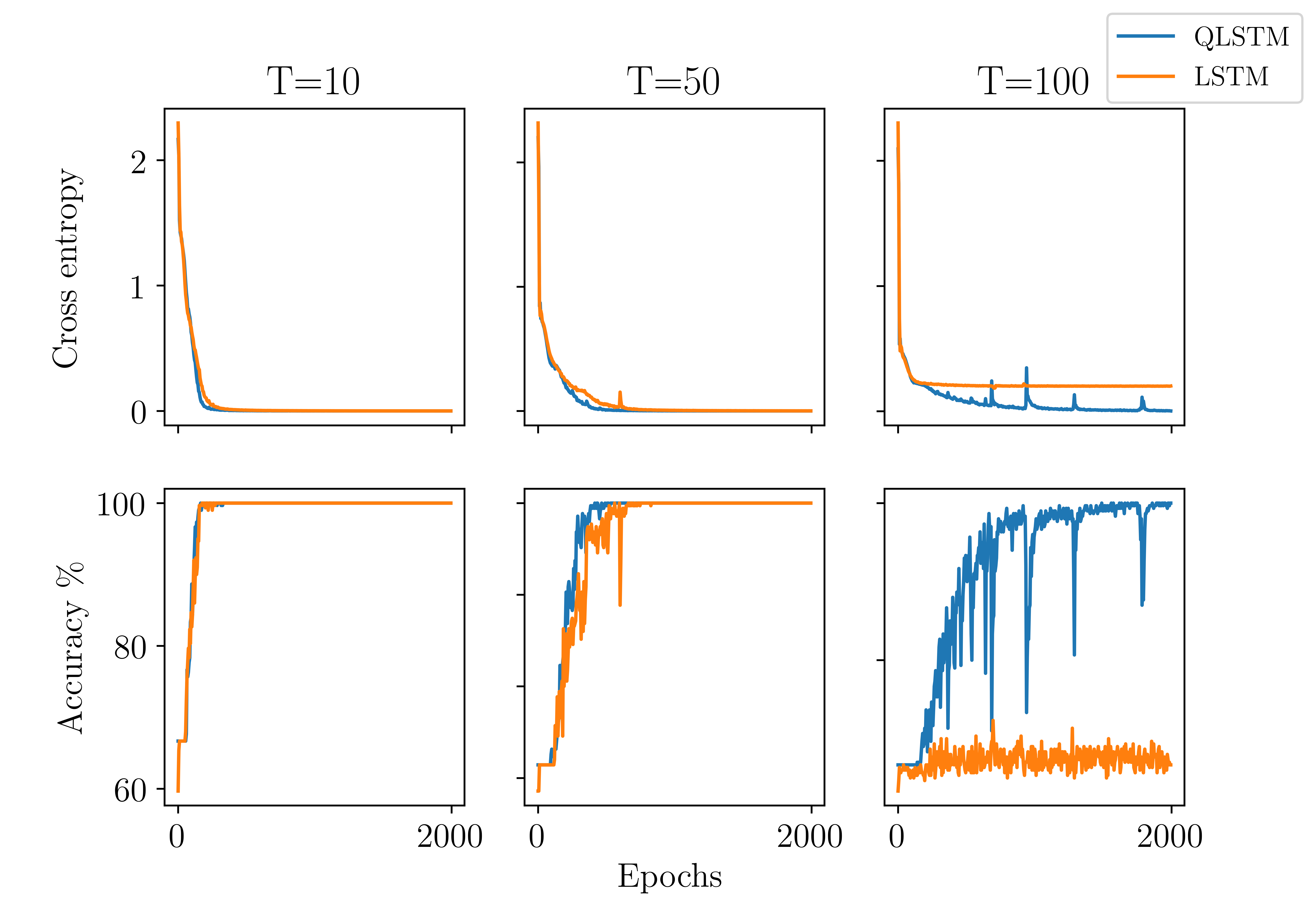}
}
\caption{Evolution of the cross entropy loss, and of the accuracy of both QLSTM (Blue curves) and LSTM (Orange curves) during the synthetic memory copy-task for time lags or \textit{blanks} $T$ of $10$, $50$ and $100$. }\label{fig:rescopy}
\end{center}
\end{figure}

The results reported in Fig.\ref{fig:rescopy} highlight a slightly faster convergence of the QLSTM over the LSTM for all sizes ($T$). It is also worth noticing that real-valued LSTM failed the copy-task with $T=100$ while QLSTM succeeded. It is easily explained by the impact of quaternion numbers during the learning process of inter-denpendencies of input features. Indeed, the QLSTM is a smaller (less parameters), but more efficient (dealing with higher dimensions) model than real-valued LSTM, resulting in a higher generalization capability: $20$ quaternion neurons are equivalent to $20\times4=80$ real-valued ones. Overall, the introduction of quaternions in LSTMs do not alter their basics properties, but it provides a higher long-term dependencies learning capability. We hypothesis that such efficiency improvements alongside with a dedicated input representation will help QLSTMs to outperform LSTMs in more realistic tasks, such as speech recognition. 

\begin{table*}[ht]
\caption{ Word error rates (WER \%) obtained with both training set (WSJ14h and WSJ81h) of the Wall Street Journal corpus. 'test-dev93' and 'test-eval92' are used as validation and testing set respectively. $L$ expresses the number of recurrent layers. Models are bidirectional. Results are from an average of three runs.}
\label{table:results}
\begin{center}
\scalebox{0.8}{
\begin{tabular}{U{3cm}U{1.85cm}U{1.85cm}U{1.85cm}U{1.85cm}U{1cm}}
    \hline\hline
    \textbf{Models} & \textbf{WSJ14 Dev.} & \textbf{WSJ14 Test} & \textbf{WSJ81 Dev.} & \textbf{WSJ81 Test} & \textbf{Params}\\
    \hline
    $\mathbb{R}$-LSTM-3L-256 & 12.7 & 8.6 & 9.5 & 6.5 & 4.0M \\
    $\mathbb{H}$-QLSTM-3L-256 & 12.8 & 8.5 & 9.4& 6.5 & 2.3M\\
    $\mathbb{R}$-LSTM-4L-256 & 12.1 & 8.3 & 9.3 & 6.4 & 4.8M\\
    $\mathbb{H}$-QLSTM-4L-256 & 11.9 & 8.0 & 9.1 & 6.2 & 2.5M\\
   	\hline
    $\mathbb{R}$-LSTM-3L-512 & \textbf{11.1} & \textbf{7.1} & 8.2 & 5.2 & 12.2M \\
    $\mathbb{H}$-QLSTM-3L-512 & \textbf{10.9} & \textbf{6.9} & 8.1 & 5.1 & 5.6M \\
    $\mathbb{R}$-LSTM-4L-512 & 11.3 & 7.0 & 8.1 & 5.0 & 15.5M \\
    $\mathbb{H}$-QLSTM-4L-512 & 11.1 & 6.8 & 8.0 & 4.9 & 6.5M\\
    \hline
    $\mathbb{R}$-LSTM-3L-1024 & 11.4 & 7.3 & 7.6 & 4.8 & 41.2M\\
    $\mathbb{H}$-QLSTM-3L-1024 & 11.0 & 6.9 & 7.4 & 4.6 & 15.5M\\
     $\mathbb{R}$-LSTM-4L-1024 & 11.2 & 7.2 & \textbf{7.4} & \textbf{4.5} & 53.7M\\
    $\mathbb{H}$-QLSTM-4L-1024 & 10.9 & 6.9 & \textbf{7.2} & \textbf{4.3} & 18.7M\\
    \hline
\end{tabular}
}
\end{center}
\label{tab:multicol}
\end{table*}

%
%
\subsection{Quaternion acoustic features}
\label{subsec:acc}

Unlike in \cite{parcollet2018qcnn}, this paper proposes to use four internal features in an input quaternion. The raw audio is first split every $10$ms with a window of $25$ms. Then $40$-dimensional log Mel-filter-bank coefficients with first, second, and third order derivatives are extracted using the \textit{pytorch-kaldi}\footnote{pytorch-kaldi is available at https://github.com/mravanelli/pytorch-kaldi} toolkit and the Kaldi s5 recipes \cite{Povey_ASRU2011}. An acoustic quaternion $Q(f,t)$ associated with a frequency band $f$ and a time-frame $t$ is formed as follows:
\begin{align}
Q(f,t) = e(f,t) + \frac{\partial e(f,t)}{\partial t}\textbf{i} + \frac{\partial^2 e(f,t)}{\partial^2 t} \textbf{j} + \frac{\partial^3 e(f,t)}{\partial^3 t} \textbf{k}.
\end{align}
$Q(f,t)$ represents multiple views of a frequency band $f$ at time frame $t$, consisting of the energy $e(f,t)$ in the filter band at frequency $f$, its first time derivative describing a slope view, its second time derivative describing a concavity view, and the third derivative describing the rate of change of the second derivative. Quaternions are used to construct latent representations of the external relations between the views characterizing the contents of frequency bands at given time intervals. Thus, the quaternion input vector length is $160/4 = 40$. Decoding is based on Kaldi \cite{Povey_ASRU2011} and weighted finite state transducers (WFST) that integrate acoustic, lexicon and language model probabilities into a single HMM-based search graph.

%
%
\subsection{Speech recognition with the Wall Street Journal}
\label{subsec:wsj}

QLSTMs and LSTMs are trained on both the $14$ hour subset ‘train-si84’, and the full $81$ hour dataset 'train-si284' of the Wall Street Journal (WSJ) corpus. The ‘test-dev93’ development set is employed for validation, while 'test-eval92' composes the testing set. It is important to notice that evaluated LSTMs and QLSTMs are bidirectionals. Architecture models vary in both number of layers and neurons. Indeed the number of recurrent layers $L$ varies from three to four, while the number of neurons $N$ is included in a gap from $256$ to $1,024$. Then, one dense layer is stacked alongside with an output dense layer. It is also worth noticing that the number of quaternion units of a QLSTM layer is $N/4$. Indeed, QLSTM neurons are four dimensional ({\it i.e.} a QLSTM layer that deals with a dimension size of $1,024$ has $1,024/4=256$ effective quaternion neurons). Models are optimized with Adam, with vanilla hyper-parameters and an initial learning rate of $5\cdot10^{-4}$. The learning rate is progressively annealed using an halving factor of $0.5$ that is applied when no performance improvement on the validation set is observed. The models are trained during $15$ epochs. All the models converged to a minimum loss, due to the annealed learning rate. Results are from a three folds average.

At first, it is important to notice that reported results on Table \ref{table:results} compare favorably with equivalent architectures \cite{graves2013hybrid} (WER of $11.7\%$ on 'test-dev93'), and are competitive with state-of-the-art and much more complex models based on better engineered features \cite{chan2015deep} (WER of $3.8\%$ with the 81 hours of training data, and on 'test-eval92').
Table \ref{table:results} shows that the proposed QLSTM always outperform real-valued LSTM on the test dataset with less neural parameters. Based on the smallest $14$ hours subset, a best WER of $6.9\%$ is reported in real conditions ({\it w.r.t} to the best validation set results) with a three layered QLSTM of size $512$, compared to $7.1\%$ for an LSTM with the same size. It is worth mentioning that a best WER of $6.8\%$ is obtained with a four layered QLSTM of size $512$, but without consideration for the validation results. Such performances are obtained with a reduction of the number of parameters of $2.2$ times, with $5.6$M parameters for the QLSTM compared to $12.2$M for the real-valued equivalent. This is easily explained by considering the content of the quaternion algebra. Indeed, for a fully-connected layer with $2,048$ input values and $2,048$ hidden units, a real-valued RNN has $2,048^2\approx4.2$M parameters, while, to maintain equal input and output dimensions, the quaternion equivalent has $512$ quaternions inputs and $512$ quaternion hidden units. Therefore, the number of parameters for the quaternion-valued model is $512^2\times4\approx1$M. Such a complexity reduction turns out to produce better results and have other advantages such as a smaller memory footprint while saving models on budget memory systems. This reduction allows the QLSTM to make the memory more ``compact'' and therefore, the relations between quaternion components are more robust to unseen documents from both validation and testing data-sets. This characteristic makes our QLSTM model particularly suitable for speech recognition conducted on low computational power devices like smartphones.
Both QLSTMs and LSTMs produce better results with the $81$ hours of training data. As for the smaller subset, QLSTMs always outperform LSTMs during both validation and testing phases. Indeed, a best WER of $4.3$\% is reported for a four layered QLSTM of dimension $1,024$, while the best LSTM performed at $4.5$\% with $2.9$ times more parameters, and an equivalently sized architecture.

%
%
\section{Conclusion}
\label{sec:majhead}

This paper proposes to process sequences of traditional and multidimensional acoustic features with a novel quaternion long-short term memory neural network (QLSTM). The paper introduce first a novel quaternion-valued representation of the speech signal to better handle signal sequences dependencies, and a LSTM composed with quaternions to represent in the hidden latent space inter-dependencies between quaternion features. The proposed model has been evaluated on a synthetic memory copy-task and a more realistic speech recognition task with the large Wall Street Journal (WSJ) dataset. The reported results support the initial intuitions by showing that QLSTM are more effective at learning both longer dependencies and a compact representation of multidimensional acoustic speech features by outperforming standard real-valued LSTMs on both experiments, with up to $2.8$ times less neural parameters. Therefore, and as for other quaternion-valued architectures, the intuition that the quaternion algebra of the QLSTM offers a better and more compact representation for multidimensional features, alongside with a better learning capability of feature internal dependencies through the \textit{Hamilton product}, have been validated.

\bibliographystyle{IEEEbib}

\end{document}